\documentclass[12pt,aps,prd,preprint,tightenlines,nofootinbib]{revtex4}
\usepackage{amsmath,amssymb,amsthm,amsfonts}
\usepackage{slashed}
\usepackage{graphicx}
\usepackage{subfigure}
\usepackage{dcolumn}% Align table columns on decimal point
\usepackage{hyperref}      % hypertext links %%ARXIV
\usepackage{bm}% bold math
\usepackage{epstopdf}
\usepackage{setspace}
\newcommand{\ie}{{\em i.e.}}

\newcommand{\eqsref}[2]{Eqs.~(\ref{#1}) and (\ref{#2})}

\newcommand{\figref}[1]{Fig.~\ref{fig:#1}}
\newcommand{\figsref}[2]{Figs.~\ref{fig:#1} and \ref{fig:#2}}
\newcommand{\tableref}[1]{Table~\ref{table:#1}}

\newcommand{\beq}{\begin{equation}}
\newcommand{\eeq}{\end{equation}}
\newcommand{\beqa}{\begin{eqnarray}}
\newcommand{\eeqa}{\end{eqnarray}}

\newcommand{\mscan}{m_{{\rm SCAN}}}
\newcommand{\msll}{{\hat{m}_{\tilde l l}}}
\newcommand{\msllsq}{{\hat{m}^2_{\tilde l l}}}
\newcommand{\msl}{{m_{{\tilde l}}}}

\newcommand{\pslmis}{{\hat{p}_{{\tilde l}}}}

\newcommand{\mchi}{{m_\chi}}
\newcommand{\Lab}[1]{{#1}_{\rm{Lab}}}
\newcommand{\vecpLab}[1]{\vec p_{#1~\rm{Lab}}}

\begin{document}
%\setcounter{equation}{0}
%\numberwithin{equation}{section}
\vskip1cm
\title{When a Muon Is Not a Muon--- 
Detecting Fast Long-Lived Charged Particles from
Cascade Decays Using a Mass Scan
}
\author{Iftah Galon}
\author{Yael Shadmi}
\author{Shahrazad Tarboush}
\author{Shlomit Tarem}
\affiliation{Physics Department, Technion-Israel Institute of Technology, Haifa 32000, Israel\\
\vskip1cm}

\begin{abstract}
If produced at the LHC, long-lived charged particles (LLCPs) would leave 
tracks in the muon detector. 
Time-of-Flight based methods for detecting these particles
become less efficient if the LLCPs are fast,
which would typically be the case if they are produced in the decays
of some mother particle which is either heavy or very boosted.
Thus for example, in supersymmetric models with long-lived sleptons,
the long-lived sleptons produced in neutralino decays
are often fast, with $\beta\geq0.95$ even at a 7~TeV LHC. 
We propose to use the (mis-measured) invariant mass distribution 
of ``muon''-lepton pairs, where the ``muon'' could be a slepton LLCP, 
to detect it.
This distribution peaks somewhat below the neutralino mass.
The peak can be further enhanced by evaluating the distribution
for different values of candidate ``muon'' masses.
We simulate two GMSB-like models to show that this procedure can be used
to detect the long-lived sleptons and measure both their mass and 
the neutralino mass.
\end{abstract}

\maketitle

%%%%%%%%%%%%%%%%%%%%%%%%%%%%%%%%%%%%%%%%
%%%%%%%%%%%%%%%%%%%%%%%%%%%%%%%%%%%%%%%%%%%%%
\section{Introduction}
Throughout the past four decades, collider experiments
have  unveiled the full structure of the
Standard Model (SM),
by gradually discovering  a series  of new, unstable particles
through their decays to known light particles.
It is perfectly plausible however that new particles
beyond the SM are stable 
on collider-detector time-scales, by virtue of
some new symmetry that forbids or suppresses their decays.
If such new particles are charged, they must decay on
cosmological time-scales, potentially leading to
rich and testable dark matter scenarios~\cite{Feng:2008ya}.
Indeed, many extensions of the SM predict Long-Lived
Charged Particles (LLCPs) that would not decay in the LHC detectors.
These include for example sleptons in gauge-mediated 
or gravity-mediated supersymmetry-breaking models with a light gravitino,
or KK-leptons in extra-dimension models (see for example the 
review~\cite{Fairbairn:2006gg} and references therein).

If produced at the LHC, an LLCP would traverse the entire detector,
much like a muon, but unlike a muon, its speed $\beta$ would be smaller
than one.
There has been significant progress in  recent
years on (i) modifying trigger algorithms to ensure that
LLCPs are not missed altogether if they are very slow,
(ii) developing techniques for LLCP 
detection~\cite{Nisati:1997gb,Connolly:1999dv,Tarem:2009zz,
Aad:2011yf,Aad:2011hz,Khachatryan:2011ts}.
The latter, which have already led to the exclusion 
 of various LLCP models~\cite{Khachatryan:2011ts,Aad:2011yf,Aad:2011hz}, 
mostly rely on the low LLCP speed
in order to distinguish it from a muon.
At the LHC, LLCPs could be produced either directly,
or in cascade decays of heavier particles\footnote{Thus for example, 
long-lived sleptons could come from Drell-Yan
pair production, from cascade decays of strongly interacting
particles, or (if the masses of these strongly interacting
particles are very large), from the electroweak production of
gauginos, followed by the decay of these gauginos to lighter sleptons.}.
A large fraction of the latter may have $\beta$ close to 
one\footnote{This would be even more of a problem if the LHC eventually goes
to 14~TeV.}. These fast sleptons will be hard to distinguish from muons
based on Time-of-Flight methods.
Fortunately, the fact that such fast LLCPs typically
originate from the decay of a heavy particle, can provide new
handles on them.
Consider a long-lived slepton $\tilde l$ which is produced in association
with a lepton $l$ from a neutralino decay, 
$\tilde \chi^0 \to \tilde l^\pm l^\mp$. If the slepton is fast,
it can be mistaken for a muon, so that its 4-momentum is not measured
correctly. The (mis)measured invariant mass of the
opposite-sign (OS) lepton-slepton pair is therefore different from
 the neutralino mass.
Still, if the fast slepton originated from a boosted neutralino,
its mass is small compared to all the energy scales characterizing the
decay,
so that the mis-measured invariant mass distribution
would peak somewhat below $\mchi^2-\msl^2$.
This peak may be observable above the background.
Furthermore, the peak can be enhanced by scanning over the possible slepton
masses in order to obtain the correct neutralino peak.
Taking all the OS ``muon''-lepton pairs 
(where the ``muon'' could be either a real muon or a slepton LLCP), 
we can assign the ``muon'' a trial mass $\mscan$,
and calculate the resulting ``muon''-lepton invariant mass.
Varying $\mscan$, we will find that the highest peak 
in the ``muon''-lepton distribution is obtained 
when $\mscan$ coincides with the true slepton mass.
If this peak is indeed observable over the background,
one can isolate the LLCP candidates. In the process,
both the slepton and neutralino masses are measured, the first
from the best value of $\mscan$, and the second from
the position of the invariant mass peak, calculated with the
correct slepton mass.
Finally, since at the end of the process the slepton and lepton 4-momenta
are fully measured, one can hope to obtain information about
the slepton and neutralino couplings and spin, as well
as reconstruct the full event~\cite{Feng:2009bd}.

The applicability of our methods depends of course on the
flavor of the LLCP. 
As a first step, we will study here a GMSB-like example
with degenerate, long-lived right-handed sleptons,
and consider only ``muon''-electron pairs, which have little
SM background in comparison with di-muon pairs.
On the other hand, compared to models in which a single slepton
is meta-stable, our model results in a factor of three reduction 
in the efficiency for sleptons. Furthermore, uncorrelated
stau-electron and smuon-electron pairs contribute to the
SUSY combinatorial background.
It would be interesting to extend our approach to models with 
a single LLCP slepton, whether it is a stau,
a selectron or a smuon, and we plan to investigate this in more
detail in the future.

%%%%%%%%%%%%%%%%%%%%%%%%%%%%%%%%%%%%%%%%
%%%%%%%%%%%%%%%%%%%%%%%%%%%%%%%%%%%%%%%%%%%%%
\section{The Slepton-lepton mis-measured invariant mass distribution}
When a high $\beta$ slepton goes through the muon detectors,
it can easily be mistaken for a muon.
The slepton 3-momentum is then measured correctly,
but the energy is computed from this 3-momentum
assuming a zero slepton mass.
Consider then such a mis-measured slepton,
which originates from the neutralino decay
$\tilde \chi^0 \to \tilde l^\pm l^\mp$.
The measured, or rather, mis-measured,
$\tilde l l$ invariant mass is, 
\beq\label{mslldef}
\msllsq = (p_l +\pslmis)^2\,, 
\eeq
where $p_l$ is the lepton 4-momentum and $\pslmis$ is
the mis-measured slepton 4-momentum, obtained from its true
4-momentum by setting the slepton mass to zero.

If the slepton is very fast, as would be the case if its mass
$\msl$ is much smaller than the neutralino mass $\mchi$
or the neutralino energy, one would expect the $\msll$ distribution 
to peak somewhat below the neutralino mass.
Indeed, to leading order in $1-\beta^2$, where $\beta$ is the slepton
speed in units of $c$, $\msll$ takes the simple form,
\beq\label{mslllead}
\msllsq =\left(\mchi^2-\msl^2\right)\,
\left(1-\frac{\msl^2}{\mchi^2+\msl^2}\frac{1+\cos\theta}{1-r\cos\theta}
\right)\,,
\eeq 
where
\beq
r\equiv\frac{\mchi^2-\msl^2}{\mchi^2+\msl^2}
\eeq
and $\theta$ is the angle between the slepton and neutralino direction
in the neutralino rest-frame.
We see from~\eqref{mslllead}, that $\msll$ is a non-negative
 monotonically decreasing function of $\cos\theta$ whose maximum 
is obtained when the slepton and the neutralino are collinear,
at $\msllsq \sim \mchi^2-\msl^2$ in the fast slepton approximation.
We can also invert this equation to obtain $\cos\theta$ in terms
of $\msll$, yielding the form of the $\msll$ distribution,
\beq
\frac{d\Gamma}{d\msll} \propto \frac1{(\mchi^2-\msllsq)^2}
\,
\left[1 \pm a \,\frac{\mchi^2+\msl^2}{\mchi^2-\msl^2}\,
\left(1-\frac{2\mchi^2\msl^2}{\mchi^2-\msl^2}\, \frac1{\mchi^2-\msllsq}
\right)
\right] 
\eeq
where $a\leq1$, and the sign is determined by the neutralino
polarization (if the neutralino were a scalar, one would have $a=0$).
Thus, the rate is a monotonically increasing function of $\msll$,
and peaks near its maximal value which is roughly $\mchi^2-\msl^2$.
In the Appendix, we show that this behavior persists beyond leading 
order in $1-\beta^2$,
but the maximum allowed value of $\msll$ is in general given by,
\beq\label{msllmax}
\msll^{2~\rm{max}} =
[m_{\chi}^2 - \msl^2]\left[1 - 
\frac{\msl^2}{(E_{\chi}+|\vec p_{\chi}|)^2}
\right]\,,
\eeq
which coincides with $\mchi^2-\msl^2$ in the limit of infinite
neutralino energy\footnote{Note that $\msll\geq0$ follows from the fact
that $p_l$ and $\pslmis$ are both lightlike. $\msll = 0$, however, 
can be obtained only for sufficiently energetic neutralinos,
with $E_{\chi} - |\vec p_{\chi}| < \msl$.}.  

We can thus try to use the $\msll$ distribution to detect the slepton
LLCPs. The OS ``muon''-lepton invariant mass distribution
behaves differently for ``muons'' which are true muons, and for ``muons''
which are meta-stable sleptons.
The latter have an $\msll$ distribution which is monotonically increasing
 and exhibits a peak at the maximal allowed value of $\msll$.
In contrast, the invariant mass distribution of real muon-lepton pairs
usually falls off for large values of the invariant mass, unless
of course the muon-lepton pair is correlated, and comes from the decay
of some heavy particle, such as the $Z$. Such backgrounds are however
reducible, and can be eliminated, unless of course $\mchi^2-\msl^2$
happens to lie close to the $Z$ mass.

%%%%%%%%%%%%%%%%%%%%%%%%%%%%%%%%%%%%%%%%
%%%%%%%%%%%%%%%%%%%%%%%%%%%%%%%%%%%%%%%%%%%%%
\section{The $\msll$ distribution in a GMSB-like model at a 7~TeV LHC}
In order to demonstrate our techniques, we simulate
two GMSB-like models for a 7~TeV LHC 
with the SUSY-breaking parameters
\beq
\Lambda = 5 \times 10^4 ~\rm{GeV}\,,\   
M_{\rm{msg}} = 2.5 \times 10^5 ~\rm{GeV}\,,\  
N_5 = 3 \,,\ 
\tan\beta = 5\,,\  
\rm{sign}\mu = +\,,\   
C_{\rm{grav}} = 5000\,.
\eeq
Here $\Lambda$ is the ratio of the GMSB-singlet $F$ term to the messenger
scale~\cite{Dine:1995ag,Dine:1994vc}, $N_5$ is the number of messenger 
multiplets, and
the parameter $C_{\rm{grav}}$ is the ratio of the gravitino mass
to what it would be if the only source of supersymmetry breaking
were the GMSB singlet $F$-term~\cite{ATLAS:1999fr}.
Since the latter $F$-term is typically generated from larger
$F$-terms, such large values of $C_{\rm{grav}}$ are quite plausible, resulting
in heavier gravitinos compared to the naive GMSB estimate,
and therefore in long lifetimes of the Next-to-Lightest-SuperPartner (NLSP).
For this choice of GMSB parameters, the NLSP is
a right-handed slepton\footnote{Recall that the gaugino masses grow as $N_5$,
while the scalar masses only as $\sqrt{N_5}$. Already for $N_5=3$
the neutralino is heavier than the right handed sleptons.}.
We list the superpartner masses in Table~\ref{Tab:GMSB50_spectrum}.
These masses are calculated using
SPICE~\cite{Engelhard:2009br}, which is based on
SoftSUSY~\cite{Allanach:2001kg} and SUSYHIT~\cite{Djouadi:2006bz}.
 Since we are interested in fast LLCPs, we modify this
model by taking the three right-handed sleptons to be degenerate
with a mass of either 110~GeV or 130~GeV.
\begin{table}[h!]
\begin{center}
\begin{tabular}{||c||c||c||c||}
\hline \hline 
Particle & Mass [GeV] & Particle & Mass [GeV] \\ \hline \hline
$\tilde \nu_1 $ & $319$ & $\tilde \chi^+_2$ & $480$ \\ \hline \hline
$\tilde \nu_2 $ & $319$ & $\tilde \chi^+_1$ & $348$ \\ \hline \hline
$\tilde \nu_3 $ & $319$ & $\tilde g$ & $1123$ \\ \hline \hline
$\tilde \chi^0_4 $ & $480$ & $\tilde l_{1,2,3}$ & $130/110$ \\ \hline \hline
$\tilde \chi^0_3 $ &$424$ & $\tilde l_4$ & $326$\\ \hline \hline
$\tilde \chi^0_2 $ &349$$ & $\tilde l_5$ & $328$\\ \hline \hline
$\tilde \chi^0_1 $ &$198$ & $\tilde l_6$ & $328$\\ \hline \hline
$\tilde u^1$ & $965$ & $\tilde d^1$ & $1049$\\ \hline \hline
$\tilde u^2$ & $1053$ & $\tilde d^2$ &$1051$ \\ \hline \hline
$\tilde u^3$ & $1053$ & $\tilde d^3$ & $1051$ \\ \hline \hline
$\tilde u^4$ & $1068$ & $\tilde d^4$ & $1054$\\ \hline \hline
$\tilde u^5$ & $1093$ & $\tilde d^5$ & $1096$\\ \hline \hline
$\tilde u^6$ & $1093$ & $\tilde d^6$ & $1096$\\ \hline \hline
$h^0$ &$106$ & $H^0$ & $543$\\ \hline \hline
$A^0$ &  $540$& $H^+$ &  $546$\\ \hline \hline
\end{tabular}
\end{center}
\caption{Spectrum of the GMSB model calculated from SPICE. 
The three lightest slepton masses are set by hand to 130 or 110~GeV.}
\label{Tab:GMSB50_spectrum}
\end{table}
We will refer to the two models as the 130 and 110 models, according to
the NLSP mass.

We generate a total of $10000$  SUSY events,
corresponding to an integrated luminosity 
of $146 ~\rm{fb}^{-1}$ at a $7~\rm{TeV}$ LHC. 
These come from strong-strong, strong-weak, and weak-weak
pair production, which at leading order have a cross-section of $68~\rm{fb}$. 
We do not include Drell-Yan slepton pair production, which has 
a comparable cross-section, since most of the resulting sleptons 
do not pass our selection cuts.

The dominant relevant SM background comes from 
$t\bar t$ for which we have generated a sample of $420000$ events
corresponding to a luminosity of $5 ~\rm{fb}^{-1}$ at a $7~\rm{TeV}$ LHC. 
The $t\bar t$ sample is produced at leading order with a 
cross-section of $85~\rm{pb}$.
Unless stated otherwise, all the results we show below are normalized 
to $5~\rm{fb}^{-1}$.
To generate events we used MadGraph-MadEvent (MGME)~\cite{Alwall:2007st}, 
with FeynRules~\cite{Christensen:2008py}. 
The resulting events are decayed using BRIDGE~\cite{Meade:2007js} and 
put back into MGME's Pythia-PGS 
package~\cite{MGPythia, Sjostrand:2006za, PGS} which includes
hadronization and initial and final state radiation.
The slepton and lepton momenta are then smeared according 
to reported experimental resolution\footnote{Note that we do not use
the PGS detector simulation.}.

\begin{figure}[ht]
\centering
\subfigure[~Normalized $\beta$-spectrum of sleptons for $m_{\tilde l} = 110$~GeV]{
\includegraphics[width=0.45\textwidth]{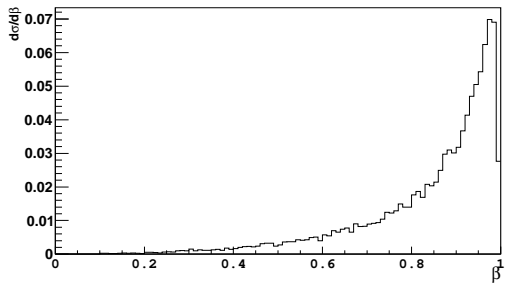}
\label{fig:beta_spectrum_110}
}
\centering
\subfigure[~Normalized $\beta$-spectrum of sleptons for $m_{\tilde l} = 130$~GeV]{\includegraphics[width=0.45\textwidth]{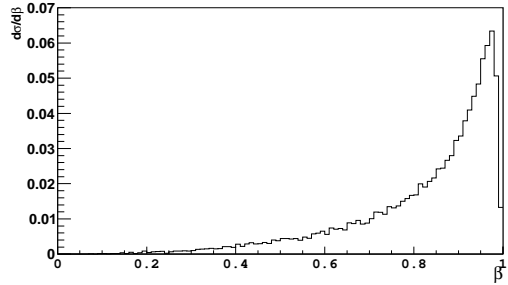}
\label{fig:beta_spectrum_130}
}
\caption{The true beta spectrum (unsmeared momenta) of sleptons.}
\label{fig:beta_spectrum}
\end{figure}

\begin{figure}[ht]
\centering
\subfigure[~$m_{\tilde l} = 110$~GeV]{
\includegraphics[height=0.3\textwidth]{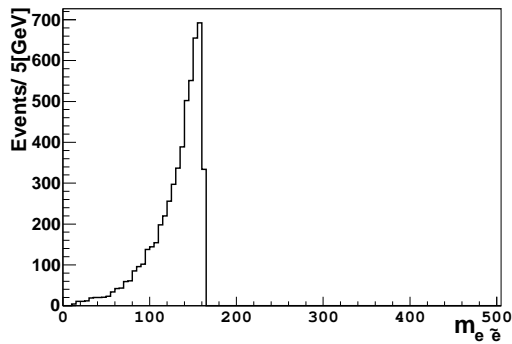}
\label{fig:110_signal_cut1}
}
\centering
\subfigure[~$m_{\tilde l} = 130$~GeV]{
\includegraphics[height=0.3\textwidth]{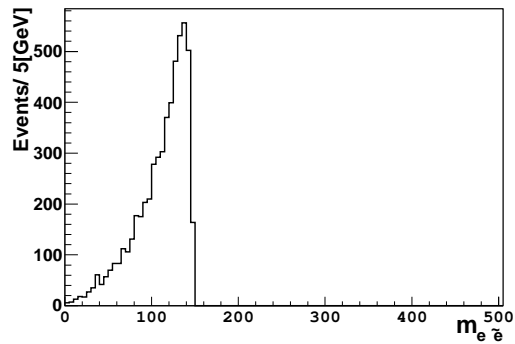}
\label{fig:130_signal_cut1}
}
\caption{The $\hat m_{\tilde e e}$ distribution of OS $\tilde e e$ pairs from the decay 
$\chi^0 \to \tilde e^\pm e^\mp$, with $p_T>20$~GeV
for both electron and selectron, for $146~\rm{fb}^{-1}$, for the two models.}
\label{fig:signal_cut1}
\end{figure}

In~\figref{beta_spectrum}
we show  the $\beta$-distributions of the sleptons in both models,
taken from truth (\ie, before any smearing).
Indeed, most of the sleptons are fast. 
For both models, roughly 50\% of the sleptons have                              
$\beta>0.9$ and 25\% have $\beta>0.95$.                                       
%%%%%%%%%%%%%%%%%%%%%%%%%%%%%%%%%%%%%%%%
%%%%%%%%%%%%%%%%%%%%%%%%%%%%%%%%%%%%%%%%
As explained in the Introduction, we are interested in fast selectrons
which are mistaken for muons. 
We propose to detect these selectrons through the selectron-lepton
mis-measured invariant-mass $\msll$. 
The $\msll$ distribution of signal events (\ie, OS $\tilde ee$ pairs 
coming from the decay $\tilde \chi^0 \to \tilde e^\pm e^\mp$)
is shown in~\figref{signal_cut1} for the two models, with a $p_T$ cut
of 20~GeV on both the electron and the selectron.
As expected, the distributions increase monotonically and peak near the
maximal allowed value of $\msllsq\sim \mchi^2-\msl^2$ (see~\eqref{msllmax}), 
which is about 150~GeV (165~GeV) for the 130 (110) model. 
%%%%%%%%%%%%%%%%%%%%%%%%%%%%%%%%%%%%%%%%
%%%%%%%%%%%%%%%%%%%%%%%%%%%%%%%%%%%%%%%%

\begin{figure}[ht]
\centering
\subfigure[~$m_{\tilde l} = 110$~GeV]{
\includegraphics[width=0.45\textwidth]{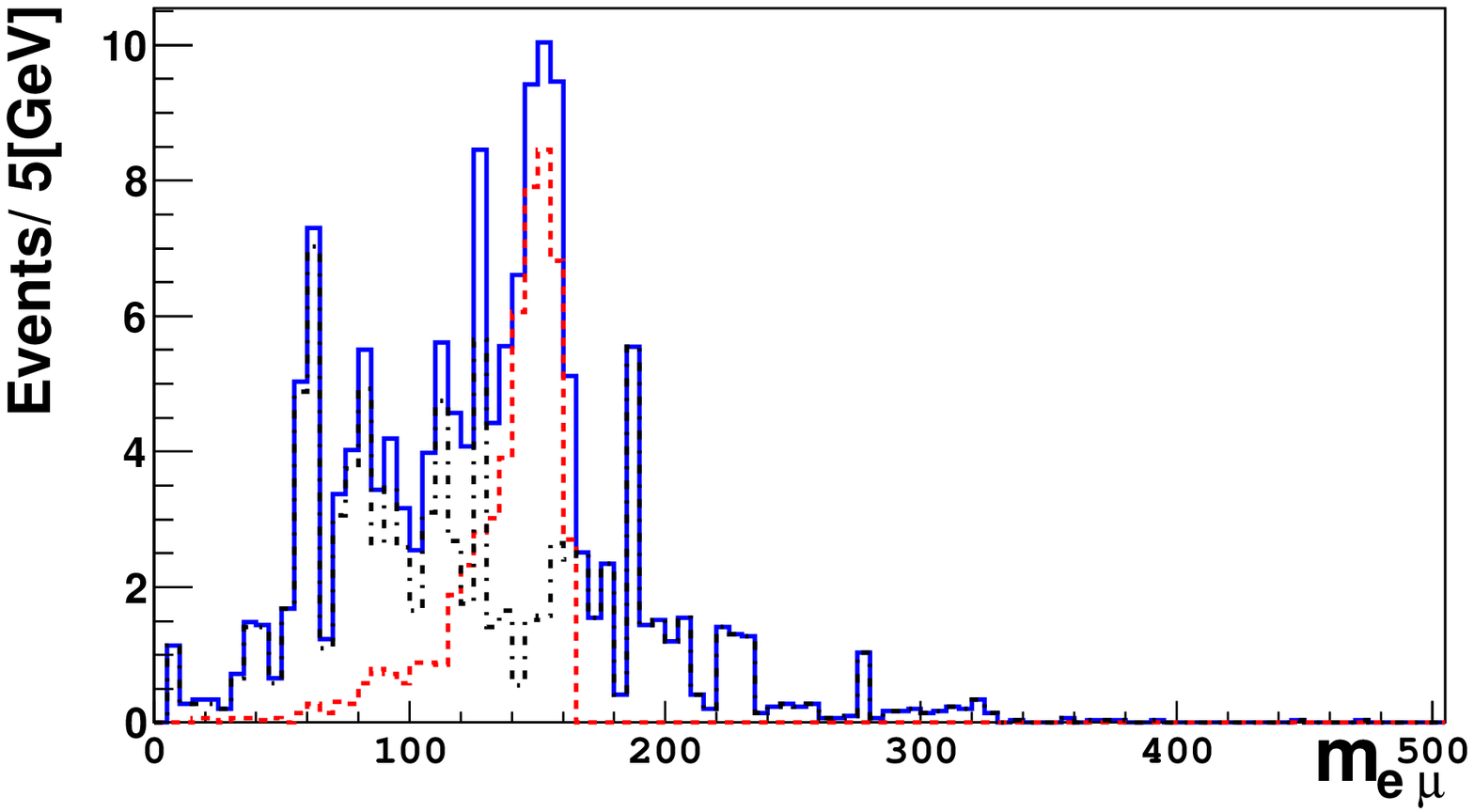}
\label{fig:110_stack_cuts123}
}
\centering
\subfigure[~$m_{\tilde l} = 130$~GeV]{
\includegraphics[width=0.45\textwidth]{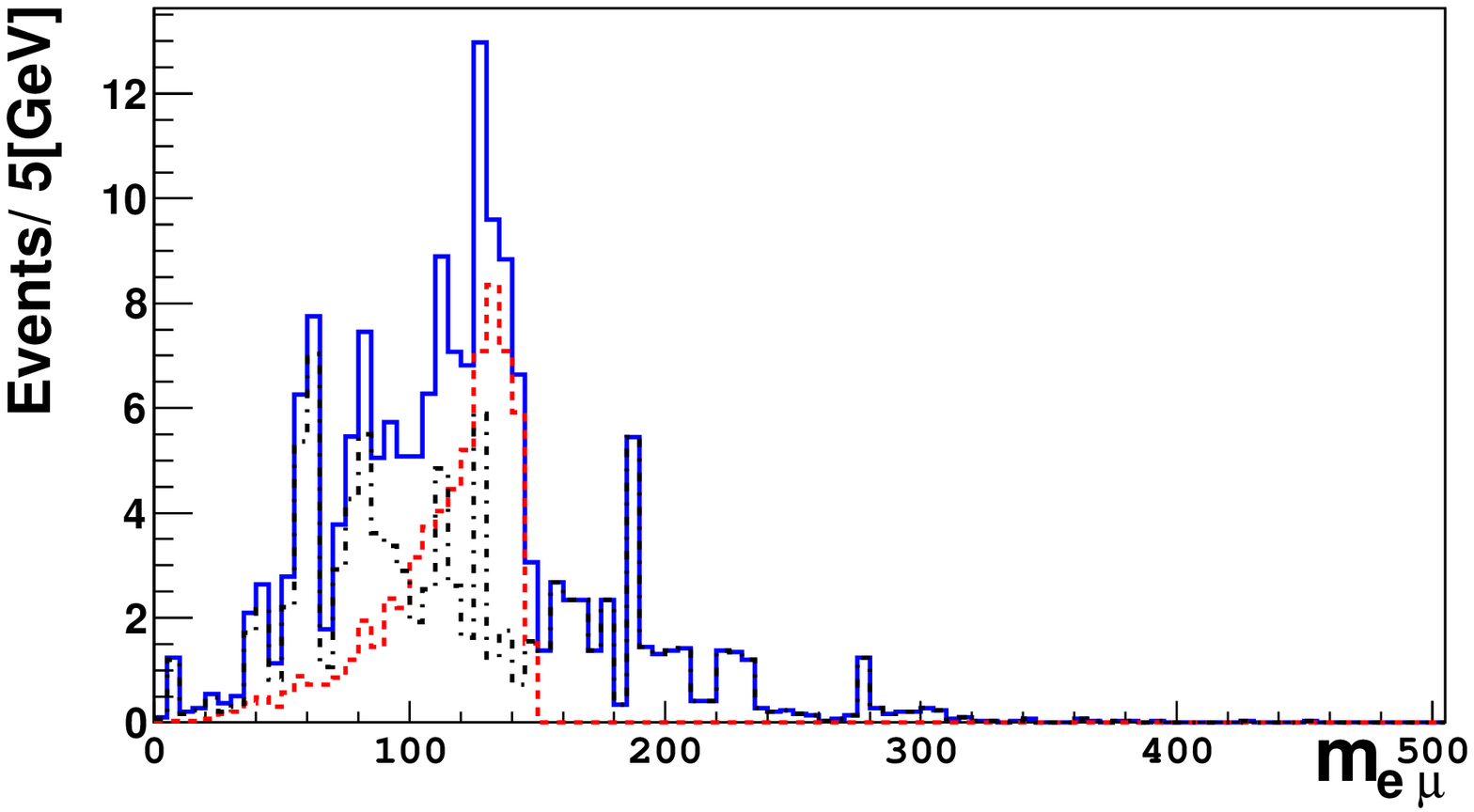}
\label{fig:130_stack_cuts123}
}
\caption{The invariant mass distributions of OS $e\mu$ pairs 
after the three cuts in the $m_{\tilde l} = 110$~GeV (left) 
and $m_{\tilde l} = 130$~GeV (right) models at $5~\rm{fb}^{-1}$.
``Muons''  in the signal distribution (red, dash) are selectrons.
``Muons''  in the background (black, dash-dot) and signal+background (blue,solid)
 distributions are either muons or any metastable-slepton 
(selectrons, smuon or stau).}
\label{fig:stack_cuts123}
\end{figure}

The relevant backgrounds include uncorrelated OS slepton-electron pairs
from SUSY events, with the sleptons in this case being either selectrons,
smuons, or staus, and OS muon-electron pairs from both SUSY events and SM
production. The main source of the latter is $t\bar t$ production.
To enhance the signal over the background we impose the following cuts:
\begin{enumerate}
\item $p_T > 20~\rm{GeV}$ on both ``muons'' and electrons. 
\item $p_T > 100$~GeV and $p > 250$~GeV on the ``muon''.
\item $\cos\theta_{\tilde l l} > 0.6$ where $\theta_{\tilde l l}$ is 
the angle between the ``muon'' and the electron.
\end{enumerate}
Here and in the following, ``muon'' denotes either a real 
muon or a slepton LLCP.
The number of events surviving these 
cuts in each of the models and in the $t\bar t$ sample 
at $5~\rm{fb}^{-1}$ are collected in \tableref{num_events}.
\begin{table}[ht]
\begin{center}
\begin{tabular}{||l||c||c||c||||c||c||c||}
\hline \hline 
	& $m_{\tilde l} = 110~\rm{GeV}$ 	&&& $m_{\tilde l} = 130~\rm{GeV}$  &&\\ \hline \hline
Sample/Cuts	& Cut 1~~~& Cuts 1+2~& Cut 1+2+3& Cut 1~~~& Cuts 1+2& Cut 1+2+3 \\ \hline \hline
Signal		&$189$	&$71$	&$53$	  &$189$	   &$81$	 &$65$	    \\ \hline \hline
SUSY BKG		&$363$	&$105$	&$30$	  &$359$        &$111$	 &$33$	    \\ \hline \hline
SUSY Sig+BKG	&$552$	&$176$	&$83$	  &$548$        &$192$	 &$98$	    \\ \hline \hline
ttbar			&$13830$	&$197$	&$70$	  &$13830$   &$197$     &$70$	    \\ \hline \hline
Total			&$14382$	&$373$	&$153$	  &$14378$   &$389$     &$168$	    \\ \hline \hline
\end{tabular}
\end{center}
\caption{The number of events surviving the set of cuts described in the text for $m_{\tilde l} = 110, 130~\rm{GeV}$ and $t\bar t$ normalized to $5~\rm{fb}^{-1}$}
\label{table:num_events}
\end{table}
The combined set of cuts significantly increases the 
signal-to-background ratio.
Still, for the models we study with $5~\rm{fb}^{-1}$, the signal peak
is not big enough to dominate  the background.
In~\figref{stack_cuts123} we show the OS $e\mu$
%%%%%%%%%%%%%%%%%%%%%%%%%%%%%%%%%%%%%%%%
invariant mass distributions: 
signal, background (SUSY background and $t\bar t$), and signal plus background
after the three cuts for the two models,
for an integrated luminosity of $5~\rm{fb}^{-1}$.
In the signal distribution  ``muons'' are selectrons, 
whereas in the two other distributions ``muons" are either real muons, 
or any metastable-slepton (selectron, smuon or stau). 
Although the signal+background distribution exhibits a small peak 
near $\mchi^2-\msl^2$, it does not clearly indicate the presence of
a decaying particle. In the next section, we will see how this
peak can be enhanced.

%%%%%%%%%%%%%%%%%%%%%%%%%%%%%%%%%%%%%%%%
%%%%%%%%%%%%%%%%%%%%%%%%%%%%%%%%%%%%%%%%%%%%%
\section{Scanning For The Slepton Mass}
\begin{figure}[h]
\centering
\subfigure[~$m_{\tilde l} = 110$~GeV]{
\includegraphics[width=0.45\textwidth]{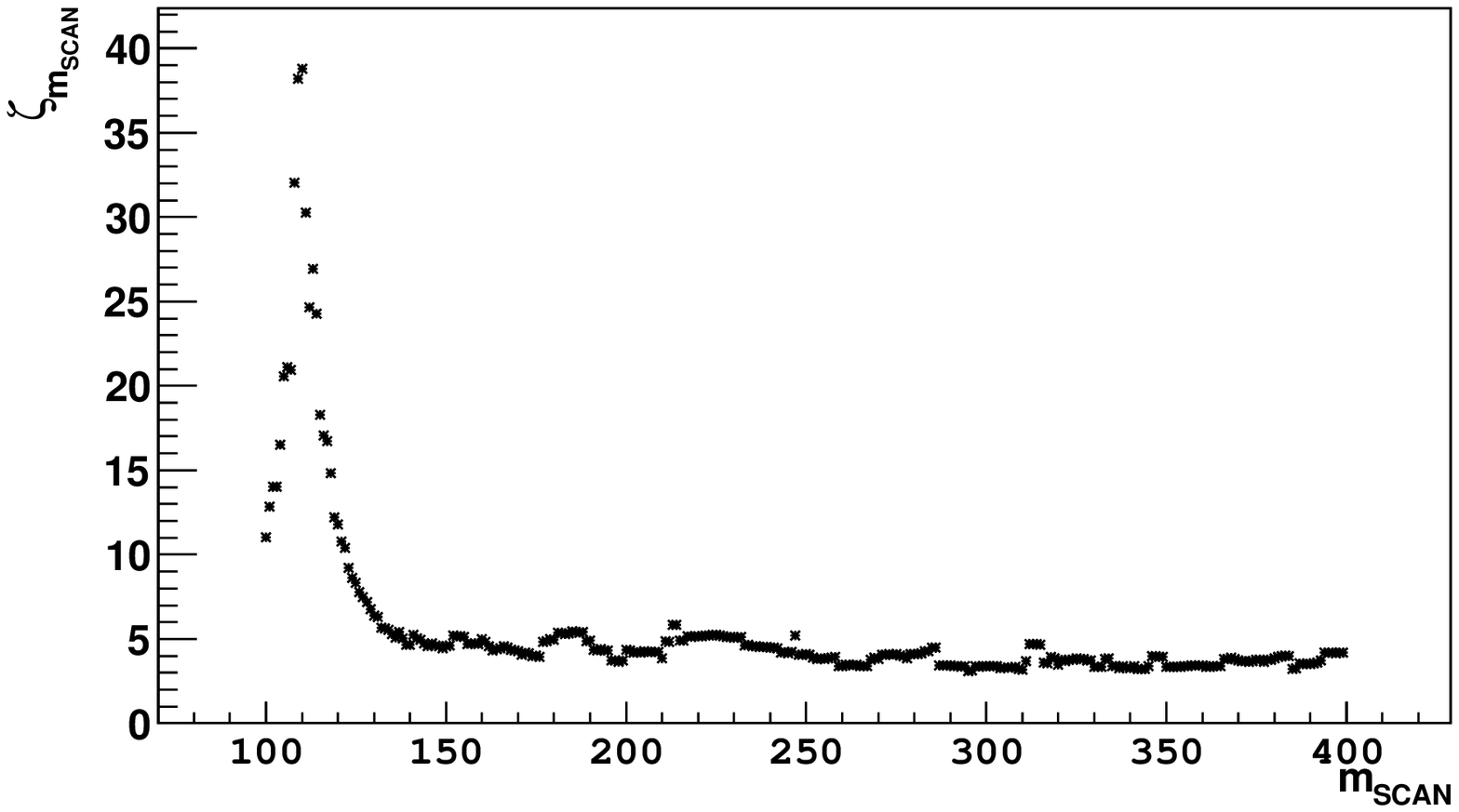}
\label{fig:Measure_110}
}
\centering
\subfigure[~$m_{\tilde l} = 130$~GeV]{
\includegraphics[width=0.45\textwidth]{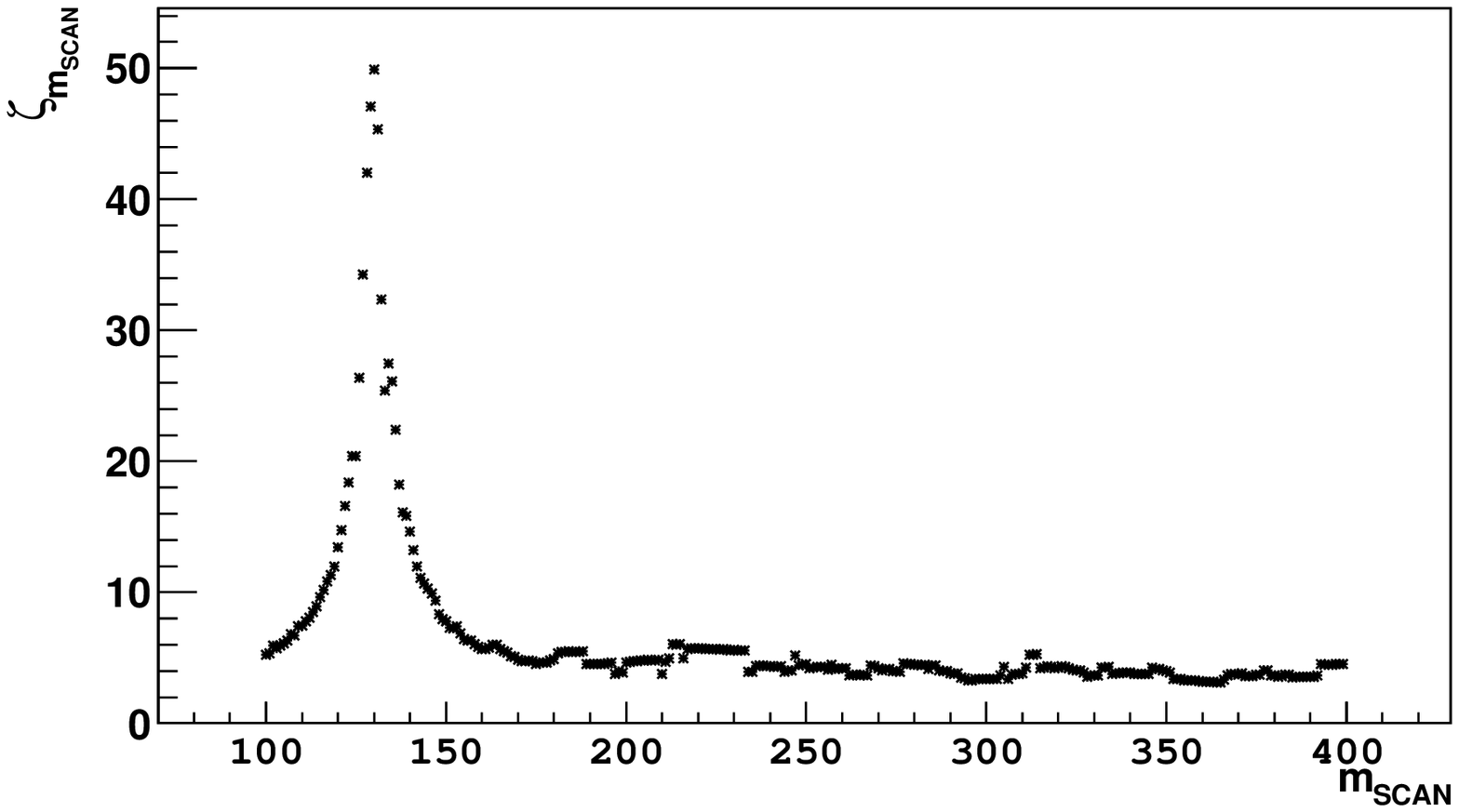}
\label{fig:Measure_130}
}
\caption{The $\zeta$ measures for each sample mass $\mscan$, for the two models.}
\label{fig:Measure}
\end{figure}

\begin{figure}[h]
\centering
\subfigure[~$m_{\tilde l}=110~\rm{GeV}$]{
\includegraphics[width=0.45\textwidth]{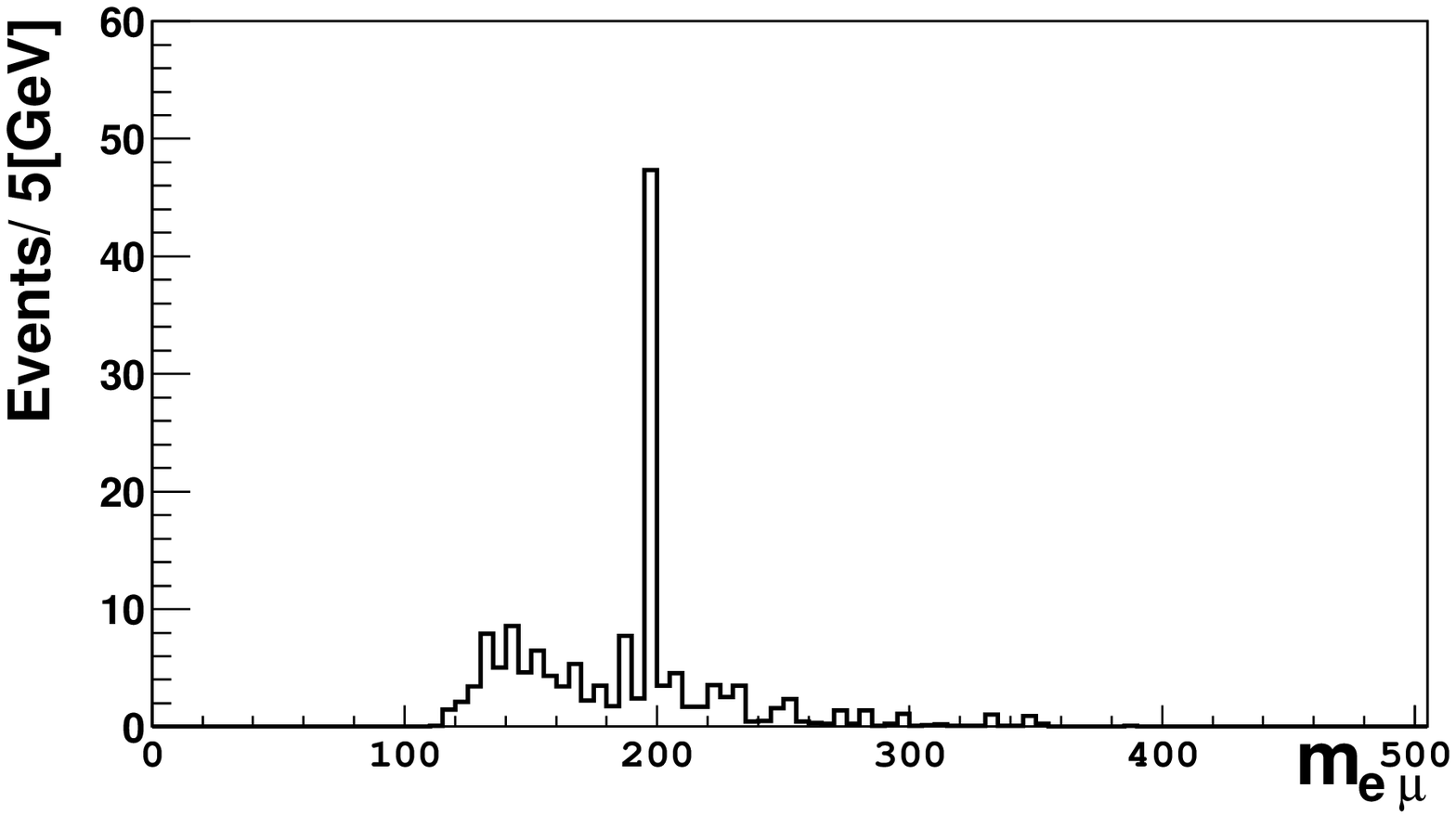}
\label{fig:110_best_mass}
}
\centering
\subfigure[~$m_{\tilde l}=130~\rm{GeV}$]{
\includegraphics[width=0.45\textwidth]{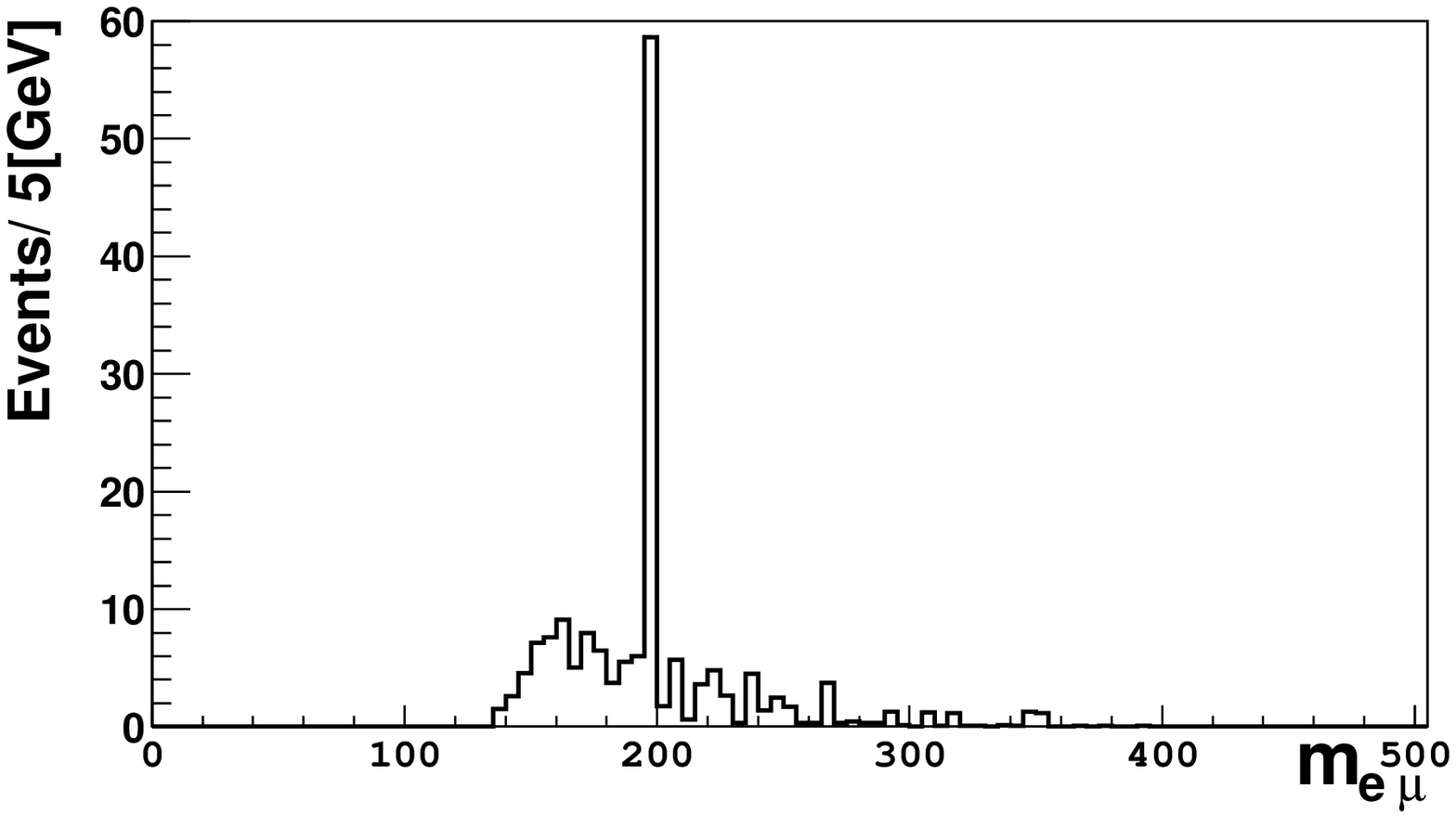}
\label{fig:130_best_mass}
}
\caption{OS $``\mu''-e$ Invariant mass distribution constructed with 
the mass $\mscan^*$  for each candidate ``muon''.
Events include both signal and background after cuts.
} 
\label{fig:best_mass}
\end{figure}

As we saw above, even after our event selection, 
the signal peak is small compared to the $e\mu$ 
background coming mainly from $t\bar t$ production. 
It is clear however how to enhance the peak: if we evaluate each
``muon'' four-momentum with the correct slepton mass, 
the signal $\msll$ distribution would peak at the neutralino
mass, while uncorrelated $\mu-e$ pairs would not.
As a result, the signal peak may be visible over the background.
We therefore study the electron-``muon'' invariant mass distributions adding
a trial mass $\mscan$ for each candidate ``muon''.
For each $\mscan$, we calculate the maximal bin height, $\zeta_{\mscan}$. 
We expect that the maximal $\zeta_{\mscan}$ would occur when $\mscan$ coincides
with the true slepton mass, since then the signal should exhibit the narrowest
peak, centered at the neutralino mass. 
In~\figref{Measure} we show $\zeta_{\mscan}$ as a function of $\mscan$,
%%%%%%%%%%%%%%%%%%%%%%%%%%%%%%%%%%%%%%%%
%%%%%%%%%%%%%%%%%%%%%%%%%%%%%%%%%%%%%%%
varying the latter over 300 values in the range 100-400~GeV, 
for the dataset obtained for the 110 and 130 models. 
Denoting the value of $\mscan$ at which $\zeta_{\mscan}$ is maximal
by $\mscan^*$, we find that in each of the two models,
$\mscan^*$ precisely reproduces the true slepton mass: 
$\mscan^*=m_{{\tilde e}}$.
We can now use $\mscan^*$ to plot the ``muon''-electron
invariant mass distribution,
calculated with $\mscan^*$  for each ``muon'' candidate.
The results are shown in~\figref{best_mass}.
%%%%%%%%%%%%%%%%%%%%%%%%%%%%%%%%%%%%%%%%
%%%%%%%%%%%%%%%%%%%%%%%%%%%%%%%%%%%%%%%%
The signal peak is clearly visible now over the background.

There are a few things we can learn from these plots.
First, the existence of a massive LLCP can be established,
and its mass  measured to be $\mscan^*$.
Second, one can now isolate a signal-rich data sample,
which includes ``muon''-electron pairs
whose $\mscan^*$-corrected invariant mass lies close to the peak.
Each ``muon'' in this sample can be re-examined, to see whether
its track can be distinguished from a true muon.
Third, we learn that the LLCPs in the signal originate from the decay
of a heavier particle. The mass of this particle is given by the
location of the peak in~\figref{best_mass}.
Finally, with this information, the signal events can be fully reconstructed,
allowing for measurements of the masses and couplings of heavier particles
with selectron and neutralino daughters.

One could worry that by evaluating the distributions
with some arbitrary value of $\mscan$, combined with
the cuts we use,  we are artificially introducing some features
into the distributions. Thus for example, for uncorrelated
OS muon pairs, with a cut of $p_2^{\rm cut}$
($p_1^{\rm cut}$) on the momentum of the slower (faster) muon,
one would expect the majority of the leptons to have momenta
close to the cut so that the peak of the $\mscan$-corrected distribution
is obtained for
\beq\label{feature}
m_{12}^{2,{\mscan}} - m_{12}^2 \sim \mscan^2 \left(1+ p_2^{\rm cut}/p_1^{\rm cut}
\right)\,,
\eeq
where $m_{12}^2$ is the true invariant mass, and $m_{12}^{2,{\mscan}}$ is
the $\mscan$-corrected invariant mass, evaluated with a mass $\mscan$
for the faster muon.

\begin{figure}[h]
\centering
\includegraphics[width=1.05\textwidth]{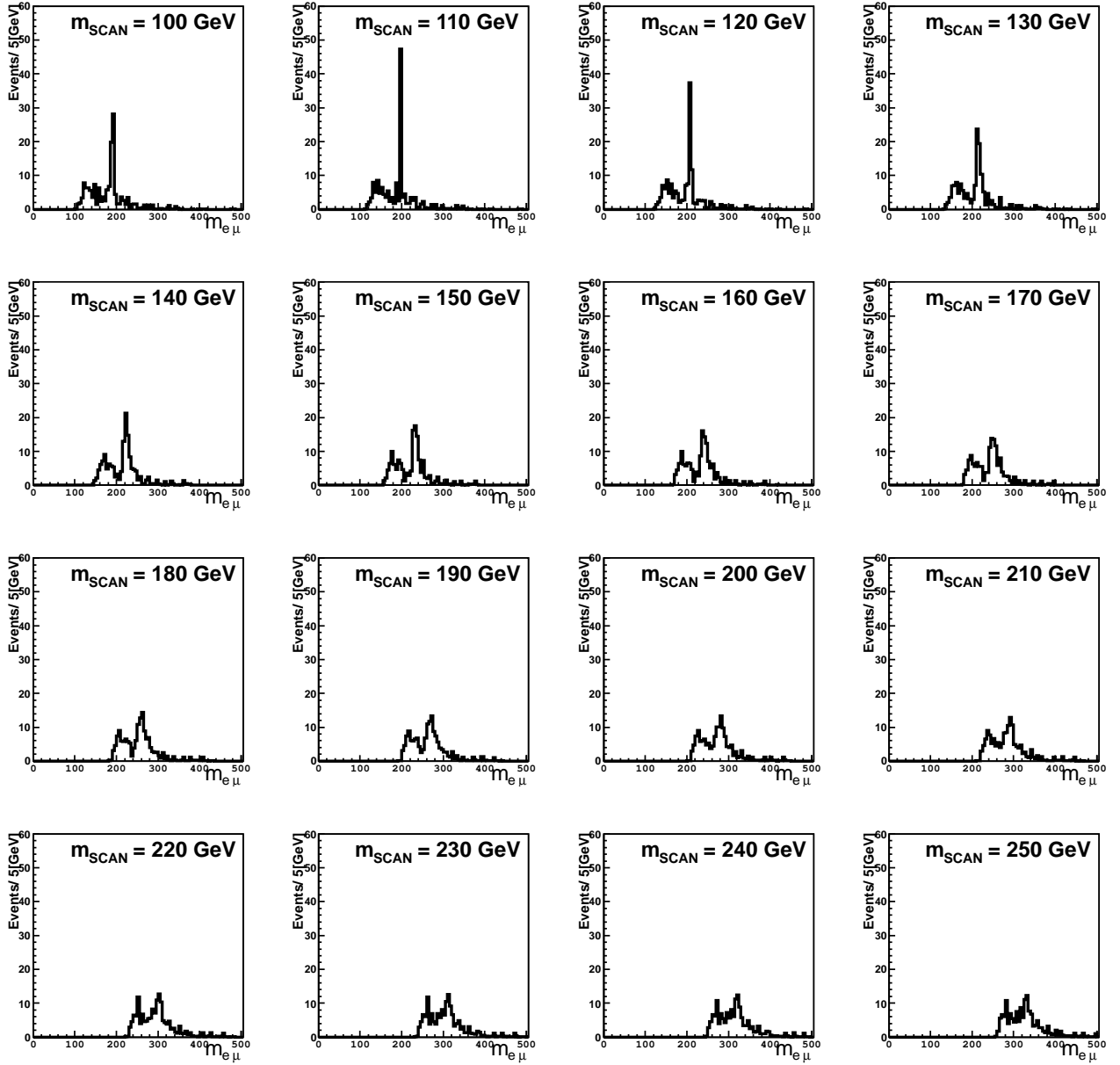}
\caption{Invariant mass distributions for each sample mass with $m_{\tilde l}=110~\rm{GeV}$} 
\label{fig:110_all_masses}
\end{figure}

To address this issue, we take two different approaches.
First, we examine the behavior of the distributions for both wrong values
of $\mscan$ and for $\mscan=\mscan^*$. 
This is done in \figsref{110_all_masses}{130_all_masses} 
%
%%%%%%%%%%%%%%%%%%%%%%%%%%%%%%%%%%%%%%%%
%%%%%%%%%%%%%%%%%%%%%%%%%%%%%%%%%%%%%%%%
\begin{figure}[h]
\centering
\includegraphics[width=1.05\textwidth]{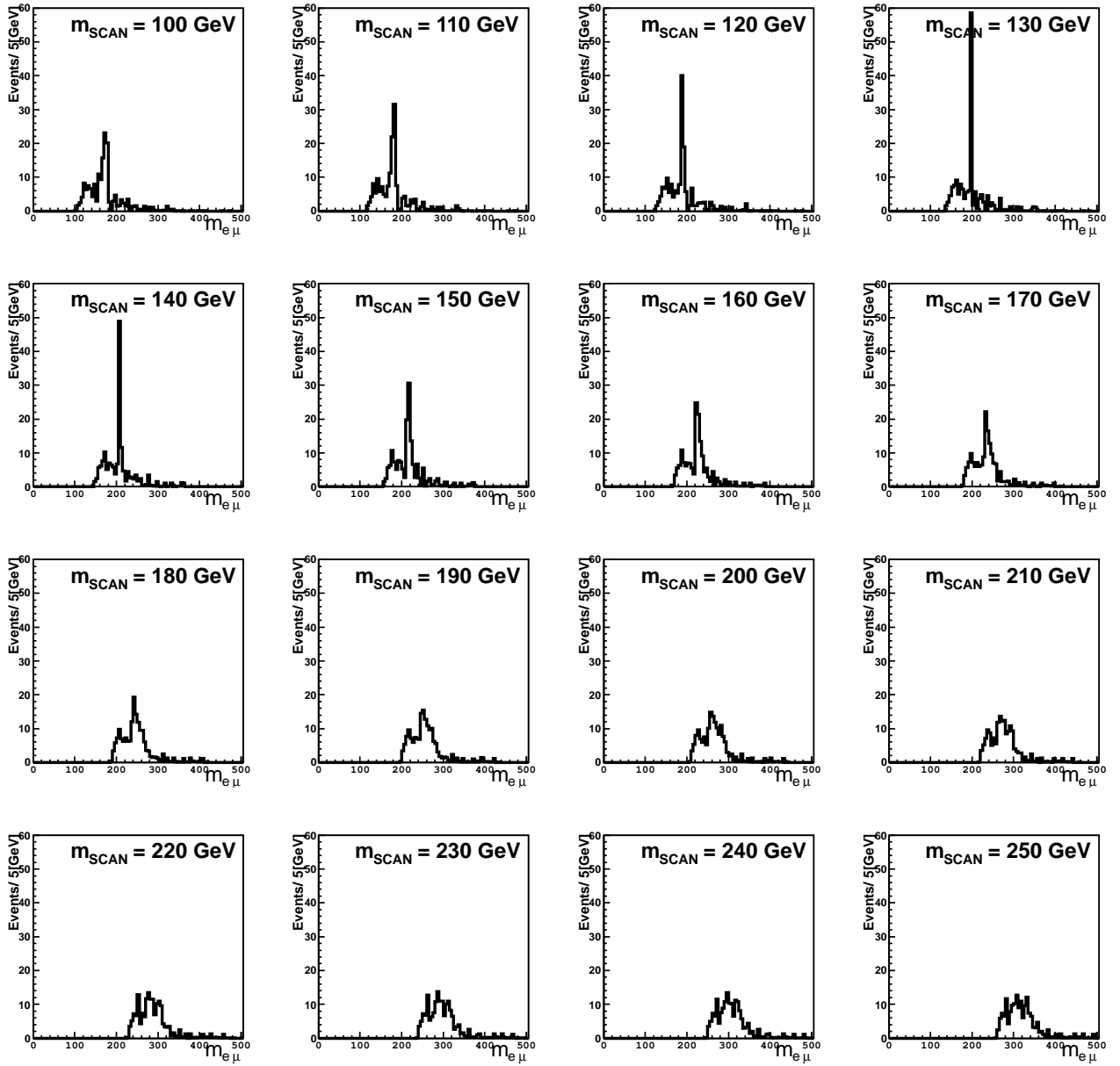}
\caption{Invariant mass distributions for each sample mass with $m_{\tilde l}=130~\rm{GeV}$} 
\label{fig:130_all_masses}
\end{figure}
%%%%%%%%%%%%%%%%%%%%%%%%%%%%%%%%%%%%%%%%
where we plot the distributions for $16$ values of $\mscan$
for each model. 
The results suggest a two-bump structure, with the lower bump
presumably corresponding to the background and lying somewhat
above $\mscan$, consistent with~\eqref{feature}.
However, for values of $\mscan$ that are very different from 
the true slepton mass no distinct features are observed. 

\begin{figure}[h]
\centering
\subfigure[~SUSY background with massless ``muon'' candidates]{
\includegraphics[width=0.45\textwidth]{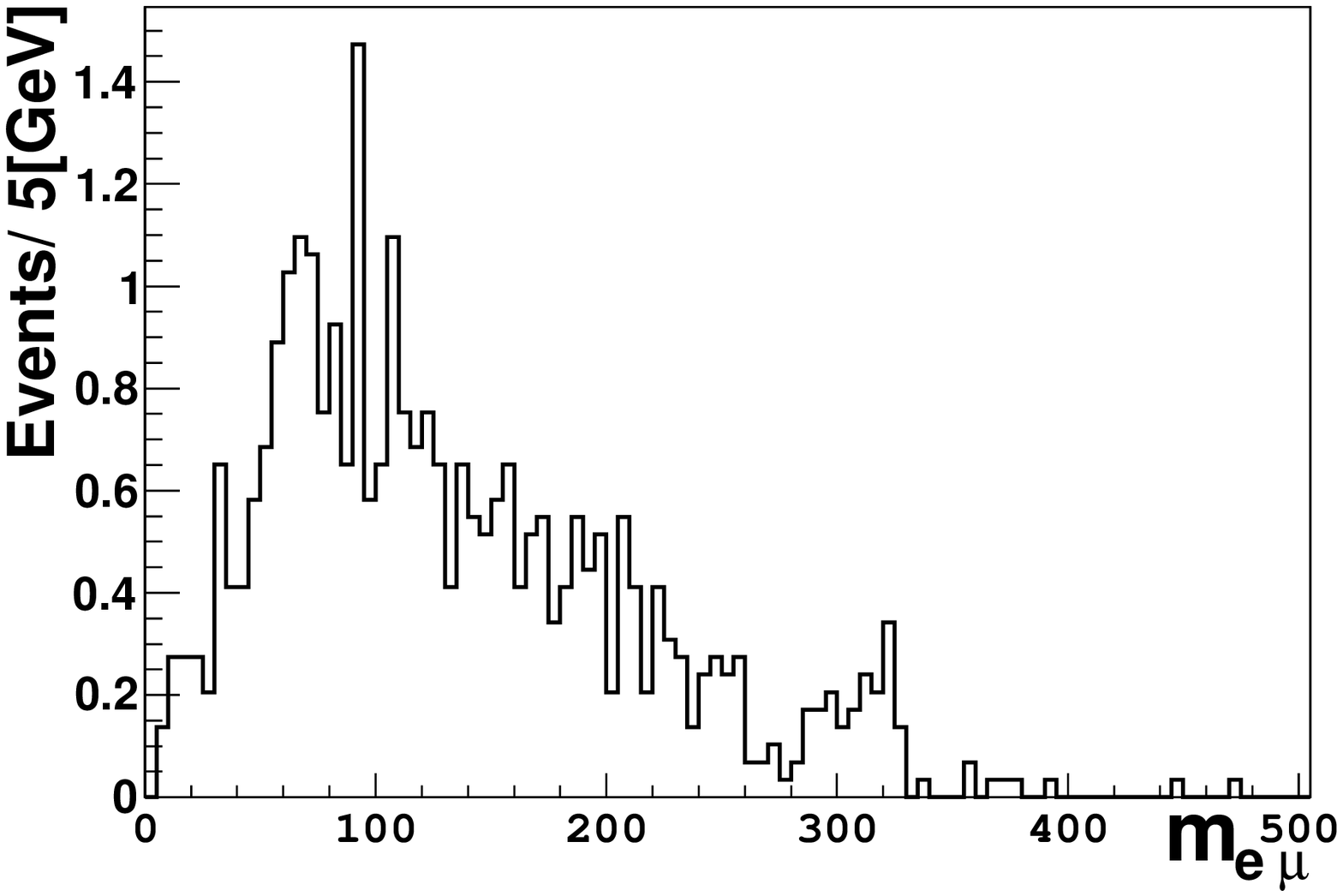}
\label{fig:110_SUSY_BKG_0}
}
\centering
\subfigure[~SUSY background  with massive ``muon'' candidates 
of mass $\mscan = 110~\rm{GeV}$]{
\includegraphics[width=0.45\textwidth]{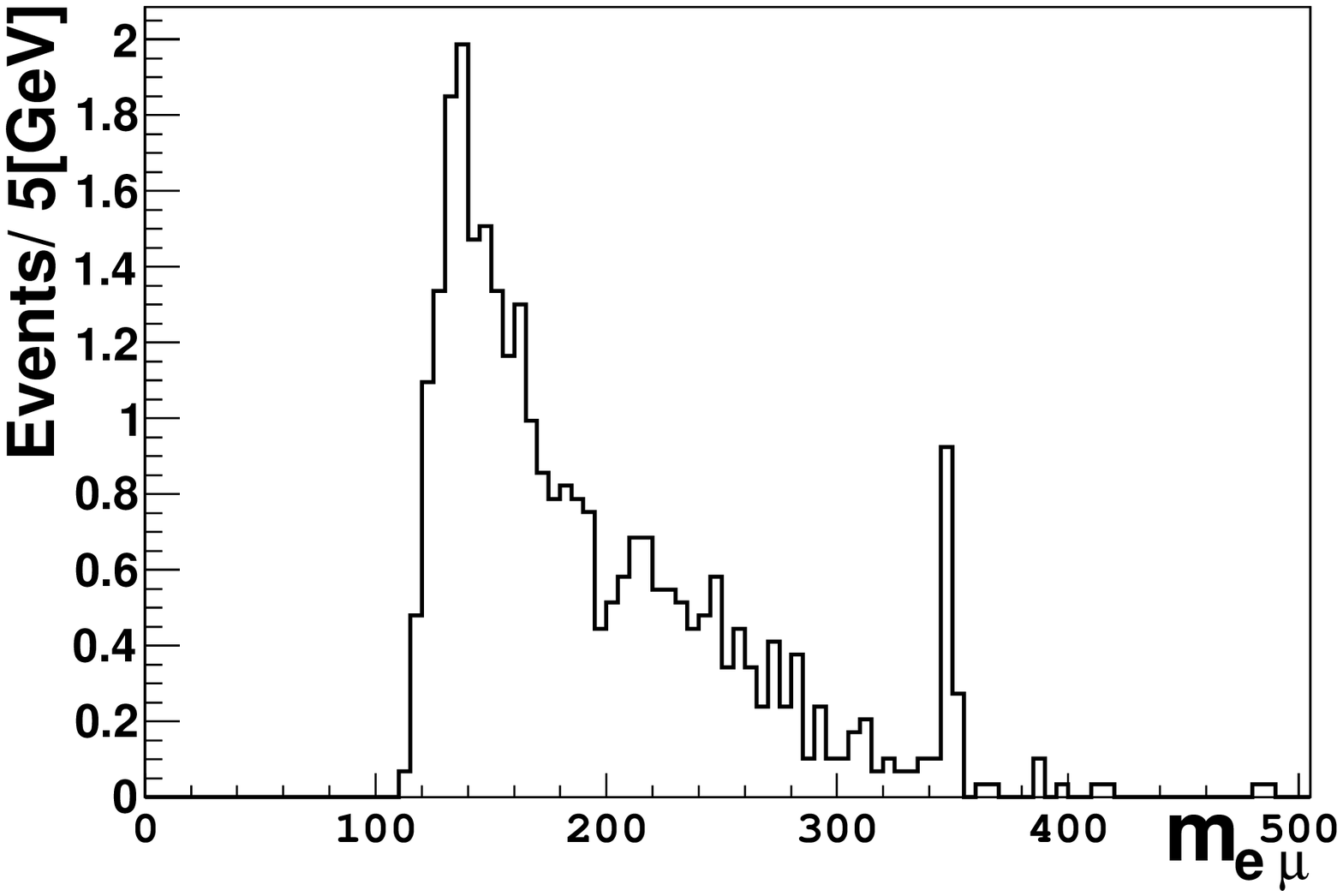}
\label{fig:110_SUSY_BKG_S}
}
\centering
\subfigure[ ~$t\bar t$ with massless muons]{
\includegraphics[width=0.45\textwidth]{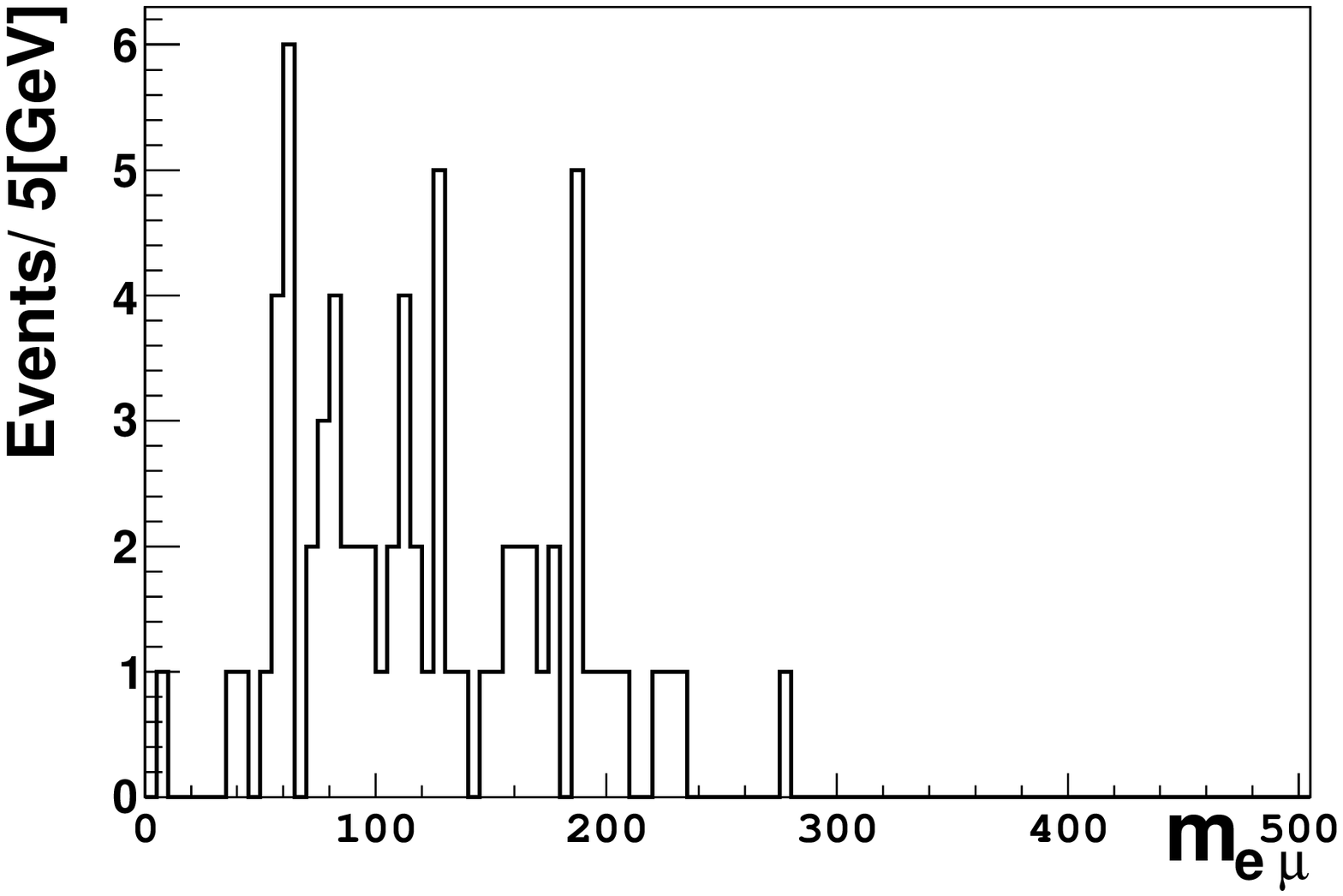}
\label{fig:110_ttbar_BKG_0}
}
\centering
\subfigure[~$t\bar t$ with massive muons  of mass $\mscan = 110~\rm{GeV}$]{
\includegraphics[width=0.45\textwidth]{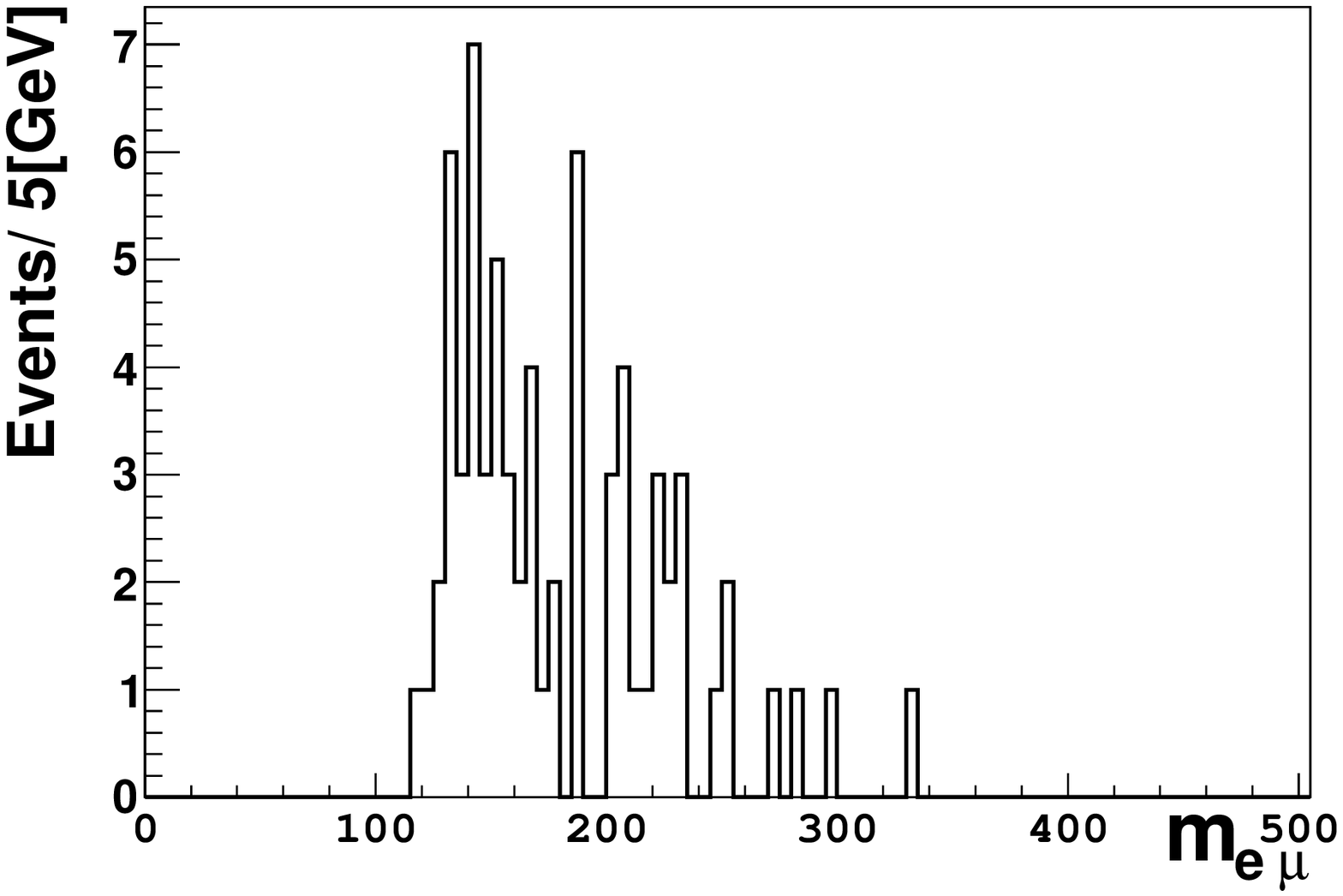}
\label{fig:110_ttbar_BKG_S}
}
\caption{The OS $e\mu$ invariant mass distribution of SUSY background (top)
in the 110 model,  
 and of the $t\bar t$ background (bottom),  after the three cuts.
The left panels show the raw distributions, and the right panels show
the distributions evaluated with $\mscan=110$~GeV.
``muons'' include both sleptons and real muons in the SUSY background 
and only real muons in the $t\bar t$ background.}
\label{fig:BKG}
\end{figure}

Second, we examine the ``raw'' and $\mscan^*$-corrected 
distributions of the SUSY background
and of the $t\bar t$ background separately from the signal 
(see~\figref{BKG})\footnote{We only show the results for
the 110 model. The results for the 130 model are very similar.} .
%%%%%%%%%%%%%%%%%%%%%%%%%%%%%%%%%%%%%%%%
%%%%%%%%%%%%%%%%%%%%%%%%%%%%%%%%%%%%%%%%%
Indeed, it is easy to see that including $\mscan^*$ does not
generate any new features in the distributions.
In fact, the one feature that does appear in the SUSY background
upon including $\mscan^*$, is a peak at around 350~GeV, which is
nothing but the $\chi_2^0$.
Due to the low SUSY cross-section, however, this ``peak'' contains
only one event.. 

Note that the cross-checks of~\figref{110_all_masses}, 
\figref{130_all_masses}, \figref{BKG}
can be done in a real experiment. This is clearly true for the first two,
but also for~\figref{BKG}. To that end, one can isolate the signal-rich sample
and the background-rich sample as described above, and plot the raw and
$\mscan^*$-corrected distribution for each one of them separately.

%%%%%%%%%%%%%%%%%%%%%%%%%%%%%%%%%%%%%%%%
%%%%%%%%%%%%%%%%%%%%%%%%%%%%%%%%%%%%%%%%%%%%%
\section{Conclusions}
We have studied a method for detecting fast long-lived sleptons
and for measuring their mass. This method relies on the
mis-measured invariant-mass distribution of slepton-lepton
pairs with the slepton faking a muon.
As a first step, we have only studied here a model with three degenerate
metastable sleptons. This has both advantages and disadvantages.
On the positive side, the model has a metastable selectron, which
results in selectron-electron pairs produced in neutralino decays.
These produce observable features in the ``muon''-electron channel
which has very low backgrounds from correlated lepton pairs.
On the down side, by using the ``muon''-electron channel only,
we are keeping only a third of the signal, compared to
models with a single metastable slepton. Furthermore, the metastable
stau and smuon can also be mistaken for muons, and therefore
contribute to the background.

The models in this work were chosen especially for their low cross sections,
so that they are not already excluded by time-of-flight or specific ionization 
methods~\cite{Aad:2011yf,Aad:2011hz,Khachatryan:2011ts}.
Though our analysis is only a ``leading-order'' analysis,
it is tempting to estimate the lower limit on the cross-sections
of detectable models.
From~\figref{best_mass} it is clear that 
much lower neutralino mass peaks (by approximately a factor of 5) 
could still be discriminated from the background. 
The efficiency of the method, however,
depends on the neutralino momentum distribution which in turn is affected
by the mass spectrum, and is thus model dependent.

The methods we describe would become even more important
in a 14~TeV LHC, since then the slepton $\beta$-spectrum 
will peak more sharply near $\beta\sim1$.

%%%%%%%%%%%%%%%%%%%%%%%%%%%%%%%%%%%%%%%%
%%%%%%%%%%%%%%%%%%%%%%%%%%%%%%%%%%%%%%%%%%%%%
\section{Acknowledgements}
The research of YS and IG was supported in part 
by the Israel Science Foundation (ISF) under grant No.~1367/11, 
and by the United States-Israel
Binational Science Foundation (BSF) under grant No.~2010221.
The work of S. Tarem and S. Tarboush is supported in part by
by the Israel Science Foundation (ISF) under grant No.~1787/11,
and by the German-Israeli Foundation (GIF) under grant 
No.~1077-94.7/2009.

\appendix

%%%%%%%%%%%%%%%%%%%%%%%%%%%%%%%%%%%%%%%%
%%%%%%%%%%%%%%%%%%%%%%%%%%%%%%%%%%%%%%%%%%%%%
\section{Mis-measured Invariant Mass Distribution in a 2-Body Decay}
\label{sec:inv_mass_mis}
In this Appendix, we will derive the functional form of
the $\msll$ distribution, and show that it is a monotonically
increasing function, and therefore peaks at the maximal
value of $\msll$. Note that, although the mis-measured
slepton momentum $\hat p^\mu$ is not a physical momentum, it is
nonetheless a 4-vector so that $\msllsq$ (see~\eqref{mslldef}) 
is Lorentz invariant
and the neutralino differential decay rate $d\Gamma/d\msll$
can be calculated in any frame.
Although it is natural to think of $\msll$ in the Lab-frame,
since this is where the slepton 3-momentum is measured,
it is instructive to express it in terms of the neutralino
energy $E_\chi$ (in the lab-frame)  and the angle $\theta$ 
between the slepton and neutralino
direction in the {\sl neutralino rest frame},
since this angle does not depend on the neutralino energy.
 
We consider the decay 
$\tilde \chi^0 \to \tilde l^\pm l^\mp$.
In the neutralino rest frame, the particle momenta are:
\begin{equation}
p^\mu |_{\rm{rest}} = (E_{\tilde l~\rm{Rest}},- \vec p_{l~\rm{Rest}}) \qquad
p_{l}^\mu |_{\rm{rest}} = (p_{l~\rm{Rest}}, \vec p_{l~\rm{Rest}}) 
\end{equation}
where $p$ denotes the slepton momentum.
and, using energy-momentum conservation, 
\begin{equation}
p_{l~\rm{Rest}} = \frac{m_{\chi}^2 - \msl^2}{2m_{\chi}}\, \qquad
E_{\tilde l~\rm{Rest}} = \frac{m_{\chi}^2 + \msl^2}{2m_{\chi}}\,.
\end{equation}
Taking the neutralino direction to be $+\hat z$,
and writing the lepton and the mis-measured slepton 4-momenta
in the Lab-frame as, 
\beq
\Lab{p_l^\mu~}  = (p_l, \vecpLab{l})\,,\ \ \ 
\Lab{\hat p^\mu~}  = (p, \vecpLab{})
\eeq
where
\beq
p \equiv |\vec p_{\rm{Lab}}| \qquad
p_l \equiv |\vec p_{l~\rm{Lab}}|\,,
\eeq
we have
\beq
\label{eq:plLab}
p_l =
\frac{p_{l~\rm{Rest}}}{m_{\chi}}\,
\left(E_{\chi} + p_{\chi}\cos\theta\right)\ .
\eeq

There are several useful ways to write
the mis-measured invariant mass-squared, 
\beq
\msll^2 \equiv
\left(\Lab{p^\mu_l}  +   \Lab{\hat p^\mu~} \right)^2
=
\left(p_l + p \right)^2 - p_{\chi}^2 = M^2-\msl^2 - 2 p_l (E-p)\,,
\label{eq:msll_def}
\eeq
where $E$ is the slepton energy, $E=\sqrt{p^2+\msl^2}$.
Note that 
$\msll^2 = \left(p_l + p \right)^2 - p_{\chi}^2$
is a function of $p_\chi$ and $\cos\theta$.

In the neutralino rest frame, we have
\beq
\label{eq:GammaRest}
\frac{d\Gamma}{d\cos\theta}
=
\frac{m_{\chi}^2 - m^2}
{32\pi m_{\chi}^3} \,|M|^2 \,,
\eeq
with
$|M|^2 = A(1\pm a\cos\theta)$,
where $a$ is a positive constant with 
$a\leq 1$, and the sign depends on the neutralino polarization.

We can write the differential decay rate in terms of $\msll$,
using \eqsref{eq:plLab}{eq:msll_def},
\begin{equation}
\frac{d\Gamma}{d\msll{ }} 
= 
\frac{|M|^2}{32\pi m_{\chi}\,
p_{\chi}} \frac{\msll{ }}{\sqrt{\msllsq+|\vec p_{\chi}|^2}}
\,\left[
\frac{\msl^2}{(E_{\chi} - \sqrt{\msllsq+|\vec p_{\chi}|^2})^2} 
- 1
\right]\,.
\end{equation}
To determine the sign of 
$d^2\Gamma/d\msll^2$ it is convenient to start from 
expression~\eqref{eq:GammaRest}, and use the chain rule to convert derivatives
with respect to $\cos\theta$ into derivatives with respect to $\msll$.
Doing that, we find that $d\Gamma/d\msll$ is a monotonic function
of $\msll$.

\providecommand{\href}[2]{#2}\begingroup\raggedright\endgroup

\end{document}